\documentclass{article}
\usepackage[T1]{fontenc} 
\usepackage[utf8]{inputenc} 
\usepackage{ismirarxiv,amsmath,cite,url}
\usepackage{graphicx}
\usepackage{color}
\usepackage{multirow}

\newcommand{\pitch}{\textup{pitch}}

\newcommand{\onsettick}{\textup{onset tick}}

\usepackage{lineno}

\title{Composer's Assistant 2: Interactive Multi-Track MIDI Infilling with Fine-Grained User Control}

\oneauthor
{Martin E.\ Malandro} {Sam Houston State University \\ {\tt malandro@shsu.edu}}

\def\authorname{M.\ E.\ Malandro}

\begin{document}

\maketitle
\begin{abstract}
We introduce Composer's Assistant 2, a system for interactive human-computer composition in the REAPER digital audio workstation. Our work upgrades the Composer's Assistant system (which performs multi-track infilling of symbolic music at the track-measure level) with a wide range of new controls to give users fine-grained control over the system's outputs. Controls introduced in this work include two types of rhythmic conditioning controls, horizontal and vertical note onset density controls, several types of pitch controls, and a rhythmic interest control. We train a T5-like transformer model to implement these controls and to serve as the backbone of our system. With these controls, we achieve a dramatic improvement in objective metrics over the original system. We also study how well our model understands the meaning of our controls, and we conduct a listening study that does not find a significant difference between real music and music composed in a co-creative fashion with our system. We release our complete system, consisting of source code, pretrained models, and REAPER scripts. 

\end{abstract}
\section{Introduction}\label{sec:introduction}

Composers using generative systems to help them create music desire the ability to steer the systems towards outputs reflective of their style and intent \cite{collab2023}. A study of the challenges that composers faced in the 2020 AI Song Writing Contest found that the systems used in that contest were not easily steerable, and called for new systems and interfaces that are more decomposable, steerable, and adaptive \cite{2020Song}. A 2023 user study of the MMM \cite{MMM, MMM2} multi-track MIDI infilling model, integrated into a digital audio workstation (DAW) with an interface containing only a temperature parameter, found that users desired additional steering control over the outputs of the model \cite{MMMeval}. Another multi-track MIDI infilling model, Composer's Assistant \cite{CA}, was adopted by a team of composers to help recreate the lost music of the opera Andromeda \cite{Andromeda}.  
Those composers expressed difficulty using the model to create melodies that fit the lyrics they already had, since lyrics tend to have a natural rhythm to them and that model does not offer rhythmic control over its outputs \cite{ManYT}.

\begin{figure}[t]
 \centerline{\framebox{
 \includegraphics[width=0.9\columnwidth, trim={0 40 80 202}, clip]{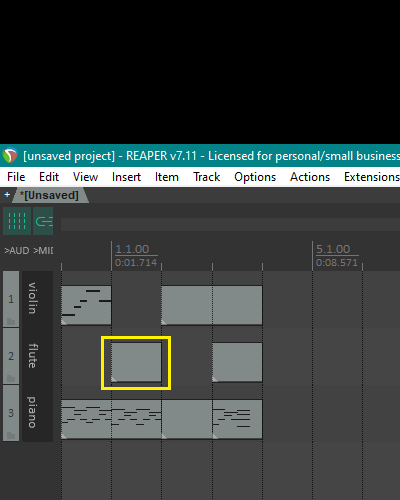}}}  

 \centerline{\framebox{
 \includegraphics[width=0.9\columnwidth, trim={0 40 80 202}, clip]{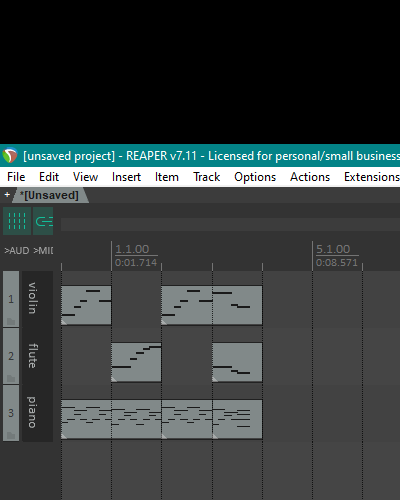}}}  

 \caption{A 4-measure prompt in REAPER, followed by a model output. Users place empty MIDI items in REAPER to tell the model in which measures to write notes, and track names to tell the model what instrument is on each track. A track-measure in the prompt is boxed. 
}
 \label{fig:prompt}
\end{figure}

Composer's Assistant (hereafter, ``CA'') is a DAW-integrated multi-track MIDI infilling model. The {\em multi-track infilling problem} is the following: Given a slice of measures from a multi-track song, where the notes have been deleted from some of the track-measures (a {\em track-measure} is a measure within a track), fill in the notes for the deleted track-measures using the notes that remain as the context---see \figref{fig:prompt}. A model trained to complete this task without any further instruction might write parts that are musically coherent but different from what the user had in mind. For instance, a composer who generates a guitar track to accompany a drum track and bass track might receive a busy, high-pitched solo, a medium-speed, medium-pitched solo, a strummed rhythmic part, or any number of other types of outputs that may not match the composer's intent for the track. It would be useful for the user to have the ability to condition the output on parameters such as rhythm (or, if a specific rhythm is not provided, horizontal note density instead), vertical note density, and pitch range.

In this work, we build upon CA to train a new model that offers a wide range of user controls. This work was guided by conversations with several composers who have used CA for co-creative composition, and represents an effort to remedy perceived shortcomings of that system. 
New controls introduced to the CA system in this work include two types of rhythmic conditioning, horizontal and vertical density controls, pitch step and leap propensity controls, several types of pitch range controls, and a rhythmic interest control. We also include a control that instructs the model not to generate octave-shifted copies of music that exists in the prompt. 
All of our controls are designed with a DAW-integrated interface in mind.
We study the power of our controls via objective metrics, we study the extent to which the model has learned the meaning of our controls, and we conduct a listening study to evaluate music created in a co-creative fashion with our model. We release our complete, DAW-integrated system and our source code.\footnote{https://github.com/m-malandro/composers-assistant-REAPER}


\section{Previous Work}
\label{sec:previous}

A wide range of generative music models predate this work, including MusicVAE \cite{musicvae}, Piano Transformer \cite{musictransformer}, Coconet \cite{coconet}, and many others \cite{musenet, symphonynet, phraseinpainting, traverselatent, drawandlisten, infill-xlnet, diffusionsymb, polyffusion}. FIGARO \cite{figaro} explored symbolic music generation with fine-grained user control, and Music SketchNet \cite{musicsketchnet} explored single-track monophonic infilling with pitch and rhythm controls.

Previous DAW-integrated models include DeepBach \cite{deepbach}, the Piano Inpainting Application \cite{pianoinpainting}, and Magenta Studio \cite{magentastudio}. Cococo \cite{cococo} is a DAW-like interface to Coconet, supporting 4 tracks and arbitrary user-driven infilling/part rewriting, similar to DeepBach. NONOTO \cite{nonoto} is a model-agnostic interface for symbolic music infilling. 


Prior multi-track infilling models include MMM \cite{MMM, MMM2}, MusIAC \cite{musiac}, and CA \cite{CA}, all of which are transformer models. The 8-bar web demo of MMM can handle up to 6 tracks, and is limited to a 4/4 time signature. MMM has two DAW-integrated versions; however, at the time of this writing they are not publicly available. MusIAC limits inputs to 3 tracks (melody, bass, and accompaniment), 16 bars, and a collection of four time signatures. CA can handle an arbitrary collection of tracks and time signatures, provided that the time signatures contain no more than 8 quarter notes per bar. 
CA is integrated into the REAPER DAW, and the underlying model runs locally on the user's machine. Each of these models offers its own set of user controls for infilling: CA offers polyphony controls, MMM offers note density and polyphony controls, and MusIAC offers five controls including note density. 

In most previous works, note density is computed simply by dividing the number of notes by the number of time steps. This means that a slow part with many thick chords can have the same density as a fast monophonic part, making it difficult for a user to steer the model towards the desired rhythmic speed with a density control. Additionally, most prior works have density control sliders (with values ranging from, e.g., 1--10) whose quantiles were defined by the training data. While this approach is attractive from a training perspective, it is difficult to navigate from a user perspective---e.g., what does a density of 7/10 mean?

\begin{figure}[t]
 \centerline{\framebox{
 \includegraphics[width=0.9\columnwidth, trim={0 40 80 202}, clip]{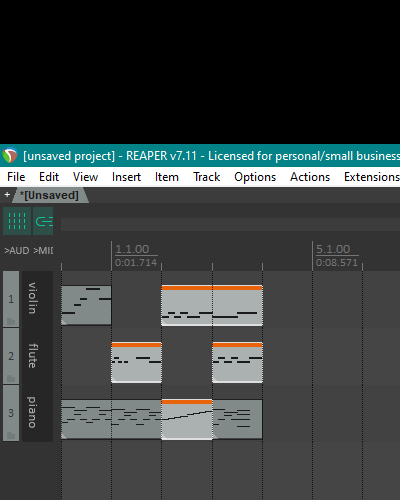}}}  

 \centerline{\framebox{
 \includegraphics[width=0.9\columnwidth, trim={0 40 80 202}, clip]{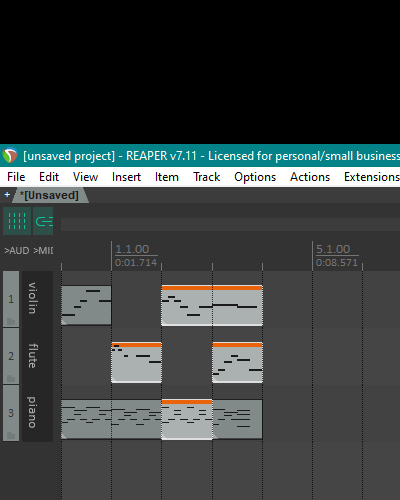}}}  

%
%
 \caption{A prompt with 1D rhythmic conditioning in REAPER, followed by a model output. Users draw the rhythms they want in the selected MIDI items, and the model chooses pitches for these rhythms that fit with the rest of the prompt. Unselected MIDI items are included in the prompt to the encoder, and remain unchanged.
}
 \label{FigRCond}
\end{figure}

In this work, we take a different approach to note density, first by factoring note density into horizontal (rhythmic) and vertical densities, and second by adopting musically meaningful quantiles for these measurements. We note that MuseMorphose \cite{musemorphose} also decomposed note density into horizontal and vertical densities, albeit with a different definition of vertical density (in their work, ``polyphony score'') than ours.
We also develop a wide range of additional steering controls, including a user option for explicit rhythmic control. With this option, the user can supply rhythms in their model prompts, and the model chooses only the pitches---see \figref{FigRCond}. 
To our knowledge, the controls we implement in this paper comprise the most comprehensive and user-friendly set of steering controls for multi-track MIDI infilling to date.




\section{Measurements for User Controls}
\label{sec:measurements}

In this section we describe the measurements underlying the user controls implemented in this work. Recall that music in MIDI format is time-quantized to a uniform grid of some number of {\em ticks} per quarter note.

\subsection{Horizontal Measurements} 
\label{SecHoriz}

\textbf{Horizontal note onset density.} 
We define the {\em horizontal note onset density} of a collection of measures from a track to be the number of ticks with a note onset divided by the total number of ticks. In interval notation, we quantize horizontal note onset densities to the following six bins: 
Less than half notes;
$[$Half notes, Quarter notes$)$;
$[$Quarter notes, Eighth notes$)$;
$[$Eighth notes, 16th notes$)$;
$[$16th notes, 4.5 onsets per quarter note$)$;
$\geq$ 4.5 onsets per quarter note.
	
\begin{figure}
 \centerline{\framebox{
 \includegraphics[width=0.9\columnwidth, trim={0 0 0 0}, clip]{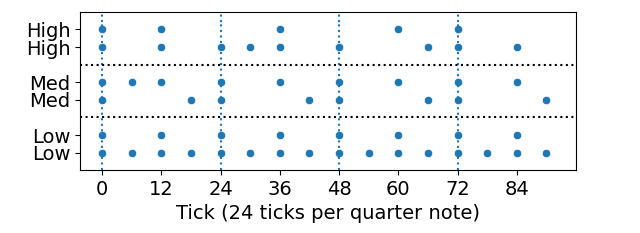}}}  

%
%
 \caption{Six examples of rhythmic interest levels. Points mark note onsets in a 4/4 measure.}
 \label{FigRhyInterest}
\end{figure}

%
	%
\medskip\noindent\textbf{Rhythmic interest.} 
Given a slice of measures from a track, let $v$ denote the binary rhythm vector of those measures. 
Let $\hat{v} = v - \bar{v}$ denote $v$, re-centered at 0. We compute dot products of $\hat{v}$ with its nontrivial shifts and record the highest of their absolute values as a measure of rhythmic uniformity. Rhythmic uniformity is scaled by $1/||\hat{v}||^2$ and subtracted from 1 to yield {\em rhythmic interest}, which we divide into Low ($<0.14$), Medium ($\geq 0.14$ and $<0.4$), and High ($\geq 0.4$) bins. These quantiles were hand-selected by looking at many examples. See \figref{FigRhyInterest} for examples.

\subsection{Vertical Measurements}

\label{SecVert}
\textbf{Vertical note onset density.} We define the {\em vertical note onset density} of a collection of track-measures from a track to be the number of notes divided by the number of ticks containing an onset. In interval notation, we quantize vertical note onset densities into the following five bins:
1 note per onset; 
(1 note per onset; 2 notes per onset];
(2 notes per onset, 3 notes per onset]; 
(3 notes per onset, 4 notes per onset]; 
$>$ 4 notes per onset.

\medskip\noindent\textbf{Average number of pitch classes per note onset.} This measure is the same as vertical note onset density, but with pitches replaced with pitch classes. It is $(\sum_{t}\# \textup{pitch classes at }t)/\#\textup{ticks containing an onset}$. We use the same bins as vertical note onset density:
1 pitch class per onset;
(1 pitch class per onset, 2 pitch classes per onset];
(2 pitch classes per onset, 3 pitch classes per onset];
(3 pitch classes per onset, 4 pitch classes per onset];
$>$ 4 pitch classes per onset.

\subsection{Pitch Measurements}
\label{SecPitch}

\textbf{Pitch step and leap propensity.} Given two consecutive notes, a {\em step} is a difference in pitch of 1--2 semitones, while a {\em leap} is a difference of more than 2 semitones. Pitch repetitions are neither steps nor leaps. We generalize to chords as follows: Given a chord $C_1$ followed by a chord $C_2$, define the chord distance $d(C_1, C_2)$ to be the average of the minimum pitch movements needed to get from the notes in $C_1$ to the notes in $C_2$:
\[
d(C_1, C_2) = \frac{1}{|C_1|}\sum_{n_1\in C_1}\min_{n_2\in C_2}|\pitch(n_1)-\pitch(n_2)|.
\]
Going from a chord $C_1$ to a chord $C_2$, we have a {\em repetition} when $d(C_1, C_2) = 0$, a {\em step} when $0 < d(C_1, C_2) \leq 2$, and a {\em leap} when $d(C_1, C_2) > 2$. Given a slice of measures from a track containing $n$ chords, we count the number of steps and leaps and divide by $n-1$ to obtain the {\em pitch step propensity} and {\em pitch leap propensity} for the slice. 
 We quantize pitch step and leap propensities into the following seven bins: $[0, 0.01)$, $[0.01, 0.2)$, $[0.2, 0.4)$, $[0.4, 0.6)$, $[0.6, 0.8)$, $[0.8, 0.99)$, $[0.99, 1.0]$.



\medskip\noindent\textbf{Note onset chromagrams.} When prompted to generate new tracks to accompany an arrangement already containing several tracks, we observed that the CA model would often generate a copy (possibly shifted by some number of octaves) of one of the tracks in the prompt. While this is often ``correct,'' it is not particularly useful for co-creative composition---if that is what the composer wanted, they could easily create this themselves. To create a control that tells the model to write genuinely new parts, for each track-measure in a song, we record whether that track-measure has the same note onset chromagram as another track-measure in its measure. (Given a track-measure $T$, the set of ordered pairs $\{(\pitch(n) (\textup{mod } 12), \onsettick(n)) : n \textup{ is a note whose onset is in }T\}$ is the note onset chromagram of $T$.) 

\subsection{Other Measurements}
\label{SecOther}

\textbf{Pitch range.} The {\em pitch range} of a collection of track measures is simply a record of their lowest and highest pitch. 

\medskip\noindent\textbf{Rhythmic information.} We define the {\em 1D rhythmic information} of a collection of measures from a track to be the set of note onset ticks and corresponding note durations after flattening all notes to the same pitch. If multiple notes share an onset, we record only the longest of their durations. The {\em 2D rhythmic information} of a collection of measures from a track is the same, but with the number of note onsets and number of pitch classes at each onset also recorded.

\section{Controls and Model}



We use the MIDI-like token-based language from \cite{CA} to tokenize music. This language supports masking of arbitrary subsets of track-measures from any slice of measures from a song. We add additional tokens to the language to serve as control tokens for the measurements introduced in \secref{sec:measurements}. For each of the measurements that are quantized into bins in Sections \ref{SecHoriz}--\ref{SecPitch}, for each bin, we create a separate control token. We also create a control token to indicate when a track-measure has a different note onset chromagram from all other track-measures in its measure---we call this token the different-note-onset-chromagram (DNOC) token. 
We also create pitch range controls and explicit rhythmic controls. For pitch range control, we create four control tokens: high (strict), low (strict), high (loose), and low (loose), and we follow such a token by a pitch token to indicate the value of the measurement. With strict pitch range controls, which can be supplied on a per-track or per-track-measure basis, the model is expected to generate at least one pitch at each extreme. With loose pitch range conditioning, the model is expected to generate at least one note within 7 semitones of each extreme and not extend beyond the extremes. The idea is that a user could supply a loose pitch range for, e.g., a vocal melody, whose bounds are given by the range of the vocalist. 
For rhythmic conditioning, we create masked pitch tokens, which are included with rhythmic tokens (describing note onset position and note duration) in prompts. For 1D rhythmic conditioning, we include a single masked pitch token at each tick containing any number of note onsets. For 2D rhythmic conditioning, we include masked pitch tokens describing both the number of note onsets and the number of pitch classes at each onset.

As in \cite{CA}, we train a T5-like \cite{T5} encoder-decoder transformer \cite{attentionisallyouneed} model. Our main model (which we refer to as our {\em large} model) is 512-dimensional, with 16 encoder layers and 16 decoder layers. This model has about 3.5$\times$ the number of parameters of the CA model, which is 384-dimensional, with 10 encoder layers and 10 decoder layers. To examine the effect of model scaling on performance, we also train a {\em small} model having the same dimension and number of layers as the CA model.
For inference, we use nucleus sampling \cite{nucleussampling} with a threshold of $p=0.85$. 

During training, we mask a random subset of track-measures from a slice of measures within a song, and we ask our model to generate the tokens for the masked track-measures. All unmasked track-measures within the slice are included in the prompt provided to the encoder. 
In each example, we include a random subset of our control tokens in our prompts. Control tokens operating on the track level are appended to the prompt, while control tokens operating on the track-measure level 
are inserted into the prompt in place of the masked tokens that the model is asked to generate. For training, values for control tokens are computed using only the masked track-measures. For inference, this allows a user to specify attributes for track-measures to be filled that differ arbitrarily from the attributes of the unmasked track-measures in the prompt. During inference in the DAW, we apply only the control tokens supplied by the user. 




We quantize music to a mixed grid that accommodates 32nd notes and 16th note triplets. This grid has 24 ticks per quarter note, of which 12 are valid locations for note onsets. 
To train our models, we use the CA training dataset (a dataset of public-domain and permissively-licensed MIDI files). 
As in \cite{anticipation}, we take the ``e'' folder of the Lakh MIDI dataset (LMD) \cite{LMD1, LMD2} to be our validation set.


\section{Evaluation}

In \cite{CA}, CA generally outperformed MMM \cite{MMM, MMM2} on objective and subjective measures. However, whether this was due to training approach, training dataset, model size, and/or other factors is unclear. 
To make a direct comparison between our work and previous work, we adopt CA as our baseline for comparison on objective metrics. 


\begin{table}[t]
\begin{center}
\begin{tabular}{|l|c|c|c|}
\hline
Model $\backslash$ Task & Random & Track & Last-bar \\
& infill & infill & fill \\
\hline
\multicolumn{4}{c}{Note $F_1$ results $\uparrow$}\\
\hline
CA2 large & \textbf{77.01}$^a$ & \textbf{70.74}$^a$ & \textbf{78.34}$^a$ \\
\hline
CA2 small & 76.27$^b$ & 69.67$^b$ & 77.24$^b$ \\
\hline
CA & 52.59$^c$ & 31.65$^c$ & 53.74$^c$ \\
\hline
\multicolumn{4}{c}{Precision $\uparrow$}\\
\hline
CA2 large & \textbf{77.15}$^a$ & \textbf{70.85}$^a$ & \textbf{78.45}$^a$ \\
\hline
CA2 small & 76.39$^b$ & 69.78$^b$ & 77.35$^b$ \\
\hline
CA & 53.02$^c$ & 33.76$^c$ & 54.72$^c$\\
\hline
\multicolumn{4}{c}{Recall $\uparrow$}\\
\hline
CA2 large & \textbf{76.90}$^a$ & \textbf{70.64}$^a$ & \textbf{78.24}$^a$\\
\hline
CA2 small & 76.15$^b$ & 69.58$^b$ & 77.15$^b$\\
\hline
CA & 52.67$^c$ & 32.22$^c$ & 53.75$^c$\\
\hline
\multicolumn{4}{c}{Pitch class histogram entropy difference $\downarrow$}\\
\hline
CA2 large & \textbf{10.78}$^a$ & \textbf{14.69}$^a$ & \textbf{8.90}$^a$\\
\hline
CA2 small & 11.28$^b$ & 15.49$^b$ & 9.75$^b$\\
\hline
CA & 31.65$^c$ & 50.53$^c$ & 31.60$^c$\\
\hline
\multicolumn{4}{c}{Groove similarity $\uparrow$}\\
\hline
CA2 large & 99.97$^b$ & \textbf{99.97}$^a$ & \textbf{99.97}$^a$\\
\hline
CA2 small & \textbf{99.98}$^a$ & \textbf{99.97}$^a$ & \textbf{99.98}$^a$\\
\hline
CA & 97.84$^c$ & 96.01$^b$ & 97.87$^b$\\
\hline
\end{tabular}
\end{center}
\caption{Objective infilling summary statistics. All cells are percentages of the form mean$^s$, where $s$ is a letter. Different letters within a metric and column indicate significant location differences ($p < 0.01$) in the samples for those table entries according to a Wilcoxon signed rank test with Holm-Bonferroni correction. 
} 
\label{TableObjectiveResults}
\end{table}

\subsection{Objective Evaluation}
\label{sec:objeval}

We take the ``f'' folder of the LMD to be our test set. From each file, we select random 8-measure slices and attempt to prepare three test examples:
\begin{itemize}
	\item Random infilling: Each track-measure in the slice is masked with probability 0.5.
	\item Track infilling: One track from the slice, containing note onsets in at least 7 measures, is masked completely.
	\item Last-bar infilling: Every track-measure in the last measure of the slice is masked.
\end{itemize}

We take 5000 examples of each type of infilling to be our test set.
For each example, we take the masked notes to be the ground truth, and after evaluating the example with our models we compute precision and recall as in \cite{CA}. The $F_1$ score for an example is the harmonic mean of its precision and recall: $F_1 = (2\cdot\textup{precision}\cdot\textup{recall})/(\textup{precision} + \textup{recall})$. We also compute the pitch class histogram entropy difference and the groove similarity (as defined in \cite{jazztransformer}) between the ground truth and the output for each example. 

We evaluate our test examples with our models, providing user controls describing the information in the masked track-measures to the maximum extent possible. In particular, we provide 2D rhythmic conditioning, pitch step and leap propensity, pitch range per track-measure, and the DNOC token (wherever applicable) to our models.
Results are averaged and presented in \tabref{TableObjectiveResults}. 
For examples evaluated by CA, we provide all of the control information available to that model---in particular, mono/poly switches at the track-measure level. We note that supplying rhythmic conditioning and pitch range information for track-measures containing only one pitch is equivalent to unmasking those track-measures. 
Therefore, to make a fair comparison between our models and CA, we unmask those track-measures in our prompts to CA and we give that model ``credit'' for those track-measures as if it had generated them itself.
(14.46\% of the track-measures from our test set fall into this category.)
We see a dramatic increase in performance of our models relative to CA. As expected, our large model outperforms our small model (except for groove similarity for the random infilling task), but the differences in performance between our models are small. Note that the groove similarity score for our models is not 100\%, indicating that our models sometimes (albeit rarely) fail to follow exactly the rhythms in their prompts. 


Next, we examine the effect of each control introduced in this paper by repeating our test examples, but with only limited control information supplied to the model. For each control introduced in this paper, we examine the $F_1$ scores obtained by supplying only that control to the model. We also examine what happens when we supply a growing collection of controls, roughly in order of how much effort is required for a user to supply such controls in practice---specifically, we supply, in order: vertical controls, horizontal controls, our DNOC control, pitch/step leap controls, pitch range (per track), 1D rhythmic conditioning, 2D rhythmic conditioning, and finally pitch range (per track-measure). The resulting large table of $F_1$ scores is omitted, but the primary observation is that pitch range and rhythmic conditioning controls have the largest positive effect on the $F_1$ scores of model outputs. Model size has only a minor effect. Horizontal, vertical, pitch step/leap propensity, and DNOC controls also have only minor effects on $F_1$ scores. This raises the questions of whether the model understands the meaning of these controls and whether these controls are useful for co-creative composition, which we address in the next two sections.

\subsection{Model Understanding of Control Tokens}

\begin{figure}[t]
 \centerline{\framebox{
 \includegraphics[width=0.9\columnwidth]{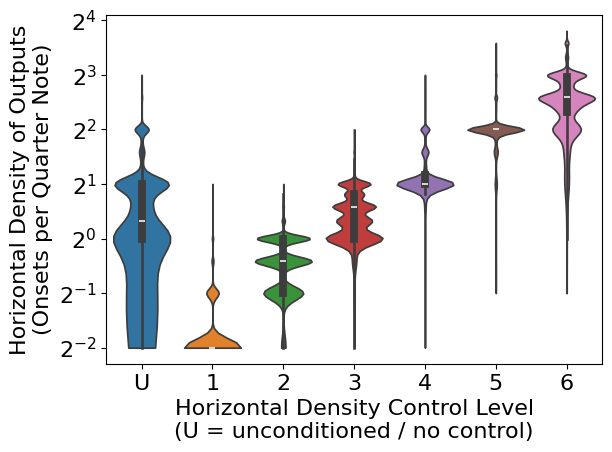}}}
 \caption{Horizontal density distributions.}
 \label{FigHorizDensity}
\end{figure}

\begin{figure}[t]
 \centerline{\framebox{
 \includegraphics[width=0.9\columnwidth]{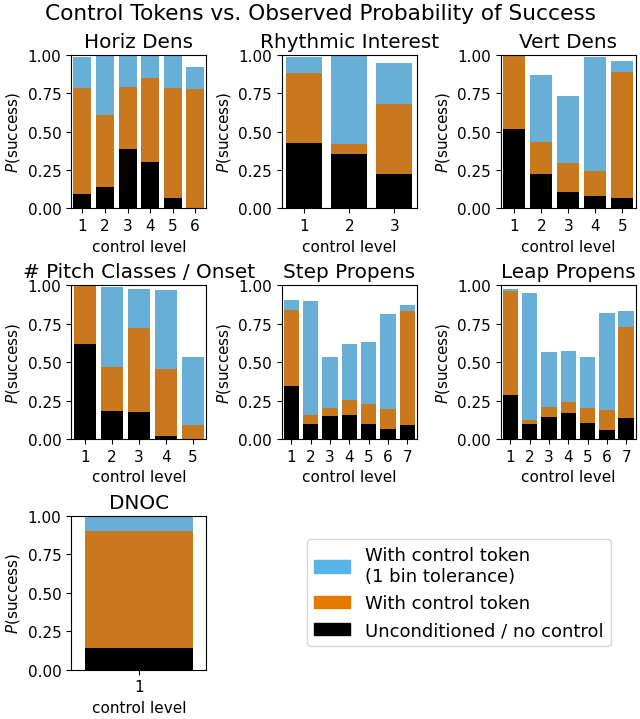}}}
 \caption{The observed probability of success for each control token in our empty prompting test.}
 \label{FigControlTokens}
\end{figure}

Our model can generate music from scratch by prompting it with prompts containing no notes. To examine the understanding our model has of our control tokens, we use an empty prompting strategy: We prompt the model to generate some number of bars of single-track music from scratch, with no control tokens, and then we repeat the empty prompt with a single control token added. We use this approach to examine our horizontal, vertical, pitch step, pitch leap, and DNOC control tokens. For each control token we generate 5000 examples, and we examine the resulting distributions. (For the DNOC control, we use a 1-bar example where a viola and cello play an ostinato one octave apart, and prompt the model to write a violin part to accompany them.)

A plot of our horizontal note onset density distributions is given in \figref{FigHorizDensity}. Plots of other distributions are similar.

For each control token, we also compare the observed probability of obtaining an output described by that token when the token is supplied versus when it is not supplied, finding that all 34 tokens mentioned above indeed push our model towards outputs having the characteristics they are intended to encode. Probabilities of success vary, however---see \figref{FigControlTokens}. Our DNOC token performs well, with a success rate of 90.5\%, as compared to unconditioned outputs having a different note onset chromagram 14.2\% of the time. Our model has also done a good job of learning the concepts of horizontal density, high and low rhythmic interest, high and low vertical density, monophonic versus polyphonic generation, and high and low step and leap propensity. However, when supplied with the appropriate control token, our model has trouble generating parts with medium rhythmic interest (with a 42\% chance of success, versus 35.2\%  for unconditioned outputs),  and with more than 4 pitch classes per onset (with a 8.92\% chance of success, versus 0.04\% for unconditioned outputs). 

Additionally, our model has not learned the precise boundaries for our bins. For instance, when asking the model to generate a part with a horizontal density in the interval $[$Quarter notes, Eighth notes$)$, we observed that our model would often generate a straight eighth note pattern. When allowing for a tolerance of up to one bin away, though, all of our control tokens have a success rate of over 50\%, and 26/34 of them have a success rate of over 80\%. 

Our model learned the meaning of each of these control tokens solely via natural training data. In future work, it may be possible to achieve better control token understanding performance by training on synthetic examples and/or by changing the boundaries of the bins.

%


\subsection{Subjective Evaluation} 
\begin{figure}
 \centerline{\framebox{
 \includegraphics[width=0.9\columnwidth]{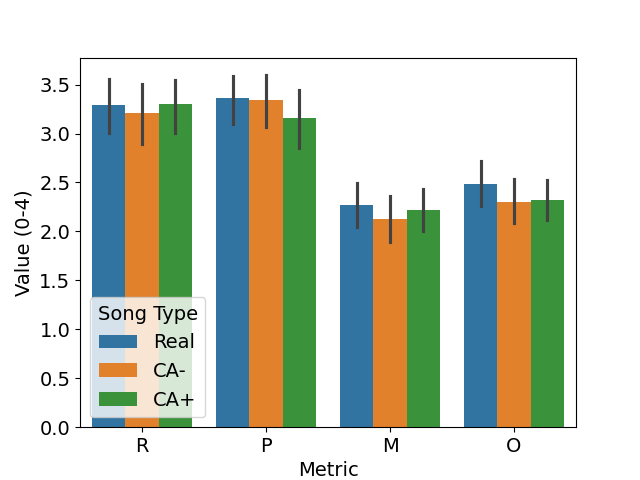}}}
 \caption{Subjective evaluation of our model.}
 \label{FigSubjectiveAvg}
\end{figure}

We used our model in an iterative, co-creative fashion to create 12 snippets of music, each about 16 bars long. To create these snippets, we started with 6 pieces of real music and deleted one or more tracks in the music. We then used our model interactively to fill these deleted tracks. Specifically, each time the model generated music, we kept the generated track-measures we liked, and continued using the model to fill the remaining track-measures until done. We carried out this  process under two sets of rules:
\begin{itemize}
	\item CA-: We did not use rhythmic conditioning, and did not hand-edit the notes in any outputs. 
	\item CA+: We were allowed to use every control available, including rhythmic conditioning, and we were allowed to make small hand edits to outputs. This is the approach closest to how the model would be used in practice.
\end{itemize}
In both cases we were allowed as many generations as we wanted. We recorded the creation of these examples, which took about 2 hours in total.\footnote{https://www.youtube.com/watch?v=WUQTlOEv3WM}
Volunteers were asked to score as many of these 18 snippets of music as they wished, according to perceived \textbf{R}hythmic correctness, \textbf{P}itch correctness, \textbf{M}emorability, and \textbf{O}verall, each on a scale of 0 to 4. Volunteers were not told which pieces of music were real and which were composed with our system. 28 individuals volunteered and ranked an average of 10.4 snippets along each of our 4 axes. Aggregated results are presented in \figref{FigSubjectiveAvg}. Each bar in this figure represents 94--99 data points.

We compare real music to music composed with our two approaches CA- and CA+ with paired $t$-tests, using data from whenever a volunteer ranked both. Each of these 8 tests involved 93--96 pairs of values. See \tabref{tab:ListeningTest}. 
After Holm-Bonferroni correction, we do not find a significant ($p < 0.05$) difference between real music and music composed with either approach in any of these 8 tests. 
We know that our model is competent at avoiding rhythmic and pitch errors, so we are not surprised by our R and P results. However, we had some difficulty creating a proper drum part in one example with the CA- approach, and we expected to find at least one significant M or O difference. We suspect that there are subjective quality differences, at least between real music and the music we created with the CA- approach, that are not large enough for this study to detect reliably. Nevertheless, we believe that this indicates that music co-created with our model has the potential to be on par with human-composed music. We also believe that results of similar studies would vary greatly depending on the skill of the composer using the system.
We invite the reader to listen to our samples\footnote{https://www.youtube.com/watch?v=4xt9fBqluQg} and to try our system. 

\begin{table}
 \begin{center}
 \begin{tabular}{|c|c|c|}
\hline
\multicolumn{2}{|c|}{Metric} & Real\\
\hline
\multirow{ 2}{*}{R} & CA- & $0.693$\\
 & CA+ & $0.887$\\
\hline
\multirow{ 2}{*}{P} & CA- &$0.791$\\ 
 & CA+ & $0.211$\\ 
\hline
 \end{tabular}
 \begin{tabular}{|c|c|c|}
\hline
\multicolumn{2}{|c|}{Metric} & Real\\
\hline
\multirow{ 2}{*}{M} & CA- & $0.052$\\ 
 & CA+ & $0.422$\\ 
\hline
\multirow{ 2}{*}{O} & CA- & $0.039$\\ 
 & CA+ & $0.099$ \\ 
\hline
 \end{tabular}
\end{center}
 \caption{Uncorrected $p$-values from paired $t$-tests in our subjective comparisons.  
}
 \label{tab:ListeningTest}
\end{table}

\section{Conclusion} 

We have introduced a wide range of steering controls for multi-track MIDI infilling, and we have trained a transformer model to implement these controls. We have created an interface for our controls, and we have integrated our model and controls into the REAPER digital audio workstation for co-creative symbolic music composition. 
Our work in this paper comprises Composer's Assistant 2, and 
we have released our complete system and source code.\footnote{https://github.com/m-malandro/composers-assistant-REAPER}

%

\section{Ethics Statement}

There are currently unresolved ethical and legal questions regarding the inclusion of copyrighted data in training sets for generative models. While we suspect it would be possible to obtain better objective (and possibly subjective) results by training a model on a larger and more varied dataset (e.g., the Lakh MIDI dataset \cite{LMD1, LMD2}), we chose to train our models only on copyright-free and permissively licensed files, primarily for the following two reasons:

First, we want our model outputs to be usable by composers. While there is a possibility that our models may output copyrighted musical information (even if such information was not present in the training dataset), we believe that training only on copyright-free and permissively-licensed musical data minimizes this possibility. 

Second, for many composers, we view the models that we have released not as the models that the composers would actually use in their work, but rather as starting points for customization and personalization. Many composers we have spoken to have said that models that write ``generic'' music are not useful to them. Instead, they want generative systems that can suggest ideas in their own style. Due to our training set, the models we have released are most proficient at infilling classical, choral, and folk music. However, informal experiments suggest that our models can be finetuned on a relatively small number of MIDI files to write in the style of those files, and hence the style limitations of our released models may not matter.
The code we have released supports finetuning by end users. While this can benefit composers who wish to use our system, there is also the risk that our models may be finetuned by users to impersonate the styles of others.

\bibliography{CA_bib}

%
%
%
%
%

\end{document}